\documentclass[a4paper,11pt]{article}
\usepackage{pos}
\usepackage{url}

\title{Theory Overview of Heavy Exotic Spectroscopy}

\author*[a]{Ciaran Hughes}
\affiliation[a]{Fermi National Accelerator Laboratory, Batavia, IL 60510, USA}
\emailAdd{chughes@fnal.gov}

\abstract{

This proceeding broadly overviews the current landscape of heavy exotic spectroscopy. Such work includes the composition of certain $X$, $Y$, and $Z$ states, and proceeds to discuss tetraquarks made exclusively of four quarks. The slides for this talk can be found at the following \href{https://indico.ipmu.jp/event/320/contributions/5130/}{link}.}

\FullConference{%
  BEAUTY2020\\
  21-24 September 2020\\
  Kashiwa, Japan (online)}

\begin{document}
\maketitle

\section{What Defines a State to be Exotic?}


In order to review exotic states, it is first necessary to know what defines a state as exotic. By definition, being exotic is being non-conventional. So, what do we mean by a conventional state\footnote{A state is rigorously defined to be a pole singularity of the S-matrix.}? A conventional state is defined to be any state which we understand ``well enough''\footnote{This definition also applies to nuclear physics, where there also exists exotic nuclei.}. Currently, mesons and baryons are the only states deemed to be conventional, since they are phenomenologically understood well enough in a potential model \cite{Brambilla:2019esw}. Everything else, such as four-quark states, are classified as exotic. This definition of conventional is ever changing, and if we understood a new class of hadrons ``well enough'', we would include this new class under the conventional umbrella. 

For illustrative purposes, the current (as of 2020) status of exotic states in the charmonium sector is shown in Fig.~1 and 2 of Richard Lebed's talk \cite{Lebed20}. Taking all exotic states from all sectors, there have been 44 experimentally observed exotic candidates, and 15 experimentally established\footnote{See the \href{https://pdg.lbl.gov/2020/reviews/contents_sports.html}{{\it{non-$q\bar{q}$ mesons}}} 2020 review from the PDG. Also, PDG defines an established signal as being seen by two independent experiments with greater than $5\sigma$ precision.} exotic candidates. Colloquially, these exotic candidates are called $XYZ$ states, but the PDG has defined a naming convention\footnote{See \href{https://pdg.lbl.gov/2020/reviews/contents_sports.html}{{\it{Naming Scheme for Hadrons}}} 2020 review from the PDG.}. 

\subsection{What Can The Exotic States Be? }

Gell-Mann knew in 1964 that color confinement allowed a plethora of states like $\bar{q}gq$, $\bar{q}\bar{q}qq$, $\bar{q}qqqq$, etc, where $q$ represents a quark, and $g$ an excited gluon degree of freedom.
For current exotic states, there are four exotic configurations largely being considered:
\begin{enumerate}
\item Hybrids. These $\bar{q}gq$ configurations exist when there is an excited gluon in an octet representation which combines with a $\bar{q}q$ in an octet representation. Phenomenological models include the constituent gluon model and the flux tube model. 
\item Molecules. A state composed of two or more constituent hadrons, typically bound by a Yakawa like force.
\item Hadro-Quarkonium. A constituent hadron core with a quark and an anti-quark cloud orbiting the core.
\item Compact Tetraquarks. A four quark state composed of a tightly arranged $qq$ diquark and a separate $\bar{q}\bar{q}$ anti-diquark. 
\end{enumerate}
The goal of exotic spectroscopy is to correctly match which exotic configurations map onto a specific experimental exotic signal. Certain models start from the proposition that exotic signals arise from a single particular exotic configuration, e.g, are solely a compact tetraquark. However, a single configuration frequently explains only certain aspects of the data, but not all. This has given rise to debate about which single configuration is most apt, for example to describe the $X(3872)$. Such debate disappears if we start from the view point that the most general solution consists of an admixture of multiple configurations. It is then not so surprising that exotic candidates exhibit complicated phenomena which can be explained by the mixture of configurations.

\section{The $\chi_{c1}(3872)$ aka $X(3872)$ }

The $X(3872)$ was the first exotic candidate discovered, in 2003 by BELLE, and consequently is the most well known and studied. Its quantum numbers have been established to be $J^{PC}=1^{++}$, $Q=0$, and $I^G=0^+$.  Its decay modes include $\mathcal{B}(\pi^+\pi^- J/\psi)> 3.2\%$. Since charmonium states like the $J/\psi$ have suppressed annihilation effects, the $X(3872)$ must have at least $\bar{c}c$ valence components. Other modes include $\mathcal{B}(D^0\bar{D^*}) > 30\%$, and $\mathcal{B}(\rho^0 J/\psi)\sim\mathcal{B}(\omega J/\psi)$. The isospin violations in this latter decay enter during the decay process, and are caused \cite{Zhou:2017txt} by the 8 MeV mass difference between the $D^0\bar{D^*}$ and $D^+\bar{D^{*-}}$ components of the $X(3872)$, i.e., isospin violations do not arise in the state itself, but during the decay process. 

One amazing piece of information about the $X(3872)$ is that its binding energy lies exactly on the $D^0\bar{D^*}$ threshold, with a binding energy of $0\pm 180$ keV. It also has a very narrow width of $\Gamma < 1.2$ MeV. Being so close to the two-particle S-wave threshold indicates cusp effects can occur, however a pure cusp explanation of the data has been ruled out \cite{Aaij:2020qga}. Both a virtual state and bound state are compatible with experimental data \cite{Aaij:2020qga}. In a potential model which only includes the $\bar{c}c$ component, the nearest state, the $\chi_{c1}(2P)$, is over 100 MeV too high \cite{Ferretti:2013faa}. However, when the $\bar{c}c$ is coupled to the two meson $D\bar{D}^*$ component, then the energy of the $\chi_{c1}(2P)$ drops and is largely in line with the experimental $X(3872)$ \cite{Ferretti:2013faa}. This is the reason that the $X(3872)$ is considered to be an exotic candidate.  Lattice QCD studies find a bound state pole with a binding energy of $\mathcal{O}(10)$ MeV at unphysical pion masses, and indicate that only $\bar{c}c$ and $D\bar{D}$ are important contributions to the wavefunction. They disfavor a diquark interpretation. Still, lattice QCD calculations with physical pion mass could find a different binding energy and pole structure.

The most likely model of the $X(3872)$ is an appreciable mixture of a $\bar{c}c$ and a $D\bar{D}$. There has been an interesting proposal that highlights that we can experimentally measure the binding energy an order of magnitude more precisely by using a triangle singularity with low energy $D^{*0}\bar{D}^{0*}$s \cite{Guo:2019qcn}.  

\section{The $\psi(4230)$ aka $Y(4230)$ aka $Y(4260)$ State}

In 2020, the PDG replaced the previously known $Y(4260)$ (found by BELLE in 2013) with the $Y(4230)$ (found by BESIII in 2017), due to the latter being more precise. The quantum numbers of $Y$ states are defined to be $J^{PC}=1^{--}$, $Q=0$, and $I^G=0^-$. The $Y(4230)$ decays to $\pi^+\pi^-J/\psi$. Because charmonium states are annihilation suppressed, the $Y(4230)$ must have a minimal $\bar{c}c$ valence component. However, the most notable statement about the $Y(4230)$ is its lack of open-charm decays into $D\bar{D}$ states - unlike excited charmonium states\footnote{Charmonium states above threshold decay into open-charm thresholds $2-3$ orders of magnitude more frequently than to hidden-charm thresholds.}. The lack of open-charm decays implies that the $Y(4230)$ has to contain more than just $\bar{c}c$, making it an exotic candidate.  

Any model for the $Y(4230)$ needs to explain the lack of open-charm decays. One model suggests that the $Y(4230)$ is a $D\bar{D}_1(2420)$ molecular state. The needed $65$ MeV binding energy is within reach of a potential model \cite{Dong:2019ofp}, making it a viable option. As molecular states decay through their constituents, this explains the $Y(4230)=D\bar{D}_1(2420)\to D\pi^+D^{*-}$ decay, as well as the $Y(4230)\to\gamma X(3872)$ decay (caused by a triangle singularity). However, by studying the $\pi\pi$ and $K\bar{K}$ distributions in the final states of $Y(4230)$ decays, \cite{Chen:2019mgp} finds that the $Y(4230)$ needs to have a sizable but not dominant $SU(3)$-flavor octet component. This means that the $Y(4230)$  needs a $SU(3)$-flavor singlet component.

The $Y(4230)$ is also consistent with a hybrid scenario. Here, lattice QCD energies of hybrids are around 180 MeV too high, but if systematic errors were included then the masses are roughly consistent \cite{Brambilla:2019esw}. Additionally, with potentials extracted from the lattice, pNRQCD finds a hybrid energy consistent with the $Y(4230)$ \cite{Brambilla:2019esw}. Due to hybrid dynamics, decays of hybrids into S-wave open charm are forbidden, making the S-P wave $D\bar{D}_1$ channel the dominant decay process. Moreover, heavy quark spin symmetry is broken more in hybrids than in quarkonia, occurring at $\mathcal{O}(\Lambda_{\text{QCD}}/m_Q)$. This could help explain the observation of the $Y(4230)$ decaying into both the heavy quark spin triplet $\pi^+\pi^- J/\psi$ and the spin singlet $\pi^+\pi^- h_c$ at comparable rates.

Given the current experimental data, the most likely model for the $Y(4230)$ is a mixture of a $D\bar{D}_1$ molecule and a hybrid, causing a bound state pole.

\section{The $Z_c(3900)$ and $Z_b(10610)$ States}
\label{sec:Z}

The $Z_b(10610)$ has decay modes $\mathcal{B}((B\bar{B}^*)^+)=86\%$, $\mathcal{B}(\Upsilon(nS)\pi^+)\sim 3\%$. For the $Z_c(3900)$ we have $\mathcal{B}((D\bar{D}^*)^{\pm}) / \mathcal{B}(J/\psi(nS)\pi^{\pm})=6.2$. Again, due to quarkonium annihilation effects being suppressed, and the $Z$ states being charged, the $Z$ states must have a minimal valence contribution consisting of four quarks $\bar{Q}Qq_1q_2$. $Z$ states are defined by the quantum numbers $Q=\pm 1,0$, $I^G=1^+$, and $J^{PC}=1^{+-}$.

For the $Z_b(10610)$, its large branching fraction into the S-wave $B\bar{B}^*$ threshold is explained by its small binding energy of $3(3)$ MeV relative to that threshold. In fact, recent lattice QCD work \cite{Prelovsek:2019ywc} extracted the potential between the $B$ and $\bar{B}^*$ and found sizable attraction in the potential for small $r$. Certain parameterisations of the potential allow for a virtual state - which would be identified as the $Z_b(10610)$. It should not be surprising then that the $Z_b(10610)$ has been identified as a virtual state when examining the experimental data by \cite{Wang:2018jlv}. As a shallow virtual state, it would be molecular \cite{Matuschek:2020gqe}, and potentially be a mixture of multiple molecular components including $B\bar{B}^*$, $\pi \Upsilon$, etc. 

Similarly, the $Z_c(3900)$ is 13 MeV lower than the S-wave $D\bar{D}^*$ threshold. A lattice QCD \cite{Brambilla:2019esw} calculation includes diquark and two meson type operators but does not find evidence of a bound state or narrow resonance. However, \cite{Albaladejo:2016jsg} indicates that this lattice work is consistent with a virtual state or broad resonance, which is also consistent with experimental data. To distinguish between these cases, smaller bin sizes and better energy resolution are needed in experimental data.

For the spatial distribution, the $Z_c(3900)$ is most likely a multi-molecular system, where the mixing between the $\pi J/\psi(\rho \eta_c) - D\bar{D}^*$ is as important as the diagonal parts of the potential \cite{Ortega:2018cnm}.

\section{Four Quark States Containing Two or More Heavy Quarks}

As we have seen, the most understood exotic states are likely a mixture of very different components, e.g., the $X(3872)$ is likely a mixture of $\bar{c}c$ and $D\bar{D}^*$. This mixing can, and does, cause complicated experimental phenomena, and this in turn makes it difficult to exclude regions in the theoretical space of models. As such, we should first attempt to find an exotic state that is solely composed of a single component and understand this state fully. Afterwards, we can use this understanding when describing the multi-component exotic candidates. 

To make full use of our intuition, we should focus on bound states rather than virtual or resonance states. A bound state is likely when at least  two of the quarks are heavy, as then there is a possibility of the formation of diquark or anti-diquark pairs due to the non-relativistic behavior. The $QQ$ diquark can be in a $\bar{3}/6$ representation with an attractive/repulsive potential. Phenomenologically, the attractive $QQ$ diquark may be deep enough in the heavy-quark potential to cause binding. For a heavy-light diquark $Qq$, there is no known mechanism derived from QCD that can cause this structure to exist, and so I will not discuss it further. 


\subsection{Bound $\bar{b}\bar{b}bb$ States}

The above reasoning prompted my coauthors and I to search for a  bound $\bar{b}\bar{b}bb$ state using lattice QCD. We searched in three channels ($J^{PC}=0^{++}$, $1^{+-}$, and $2^{++}$) and used a full basis of two-meson  and diquark anti-diquark interpolating operators. We did not find evidence for any bound state below the lowest S-wave thresholds, as shown in Fig.~9 of \cite{Hughes:2017xie}. When a signal is not found, a bound needs to be set on the likelihood that you missed that signal. This is frequently not done in lattice QCD calculations, but should be. In Fig.~11 \cite{Hughes:2017xie} we show the probability that we would have missed a bound state at a specific energy within our statistical precision. As such, we have ruled out the possibility of a $\bar{b}\bar{b}bb$ compact tetraquark  bound state to $5\sigma$. 

\subsection{Resonance $\bar{c}\bar{c}cc$ States}

In 2020, LHCb found evidence \cite{Aaij:2020fnh} of a state which decays into 2$J/\psi$. Its mass was around 700 MeV above the 2$J/\psi$ threshold, and was called the $X(6900)$. Consequently, it would be composed of $\bar{c}\bar{c}cc$. Being above threshold, this state is a resonance. A compact tetraquark with (anti-) diquark constituents is the only likely model for the state due to the non-relativistic behavior of charm quarks in this energy region. The $X(6900)$ is very exciting as it is the first clear evidence for a state that can be explained by a diquark model which can be connected to QCD.

There are a few noticeable features in Fig.~2 of the LHCb data. First, there are three bumps, and second, there is a dip around 6.8 GeV. All these features need to be explained by a model. As we have discussed, two body S-wave thresholds are important for exotic states. Some important S-wave thresholds are the $2\chi_{c0}$, $\chi_{c1}\chi{c0}$, and the $\Xi_{cc}\bar{\Xi}_{cc}$ open flavour threshold. 

As these compact tetraquark states should be narrow, the broad width of the experimental bumps hint that multiple states are contributing to the signal. LHCb have not yet performed an amplitude analysis which could distinguish nearby $J^{PC}$ states. What most models seem to agree on is that the $X(6900)$ is some combination of multiple 2S $0^{++}$ states. Then the 1S $0^{++}$ state could either be the first bump around $6.5$ GeV \cite{Karliner:2020dta}, or alternatively could be below the $2J/\psi$ threshold (where LHCb does not have data). If the 1S $0^{++}$ is below the $2J/\psi$ threshold, then the first bump would be some combination of 1P $0^{-+}$ states \cite{Giron:2020wpx}. The dip is explained by destructive interference from the $2\chi_{c0}$ threshold becoming accessible.
Further, such states are likely in the bottom sector.

\subsection{Bound $bb\bar{u}\bar{d}$ State}

A $J^P=1^+$, $I=0$ exotic state composed of $bb\bar{u}\bar{d}$ quarks has been predicted to be bound within a multitude of different theoretical frameworks. Fig.~8 of \cite{Leskovec:2019ioa} illustrates this consensus between lattice NRQCD calculations, potentials extracted from lattice QCD, and phenomenological model calculations. For a review talk of this state, and similar $Q_1Q_2\bar{q_3}\bar{q_4} $ exotic states, see \href{https://www.int.washington.edu/talks/WorkShops/int_20_2c/People/Francis_A/}{Anthony Francis' talk}. 

In fact, there is a straightforward intuitive understanding why this state is bound. First, start with the two $b$-quarks. Assume they are in the infinitely heavy quark mass limit. Then they arrange themselves into an attractive $\bar{3}$ diquark. The $\bar{u}$ and $\bar{d}$ then form a light quark cloud that screens the $bb$ interaction. This is sufficient to form deep binding around $100$ MeV. Adding in finite $b$-quark mass corrections does not change this intuitive picture. Given the robustness of this prediction, a confirmation is needed from experiment, potentially via the $bb\bar{u}\bar{d}\to \Xi^0_{bc}\bar{p}$ or $\to B^-D^+\pi^-$ processes. If found, this would be the first bound state that is composed exclusively of four quarks, and would be a very useful leverage point in our understanding of exotics.

\subsection{The Bound $D_{s0}(2317)$ State}

The $D_{s0}(2317)$ has many similar features to the $X(3872)$. First, prior to its discovery, it was expected to be the $j=\frac{1}{2}^+$ state composed of $c\bar{s}$ quarks, above threshold, and very broad through its decay to $DK$ \cite{Ferretti:2013faa}. However, experimentally, the $D_{s0}(2317)$ state is below the $D^0K^+$ threshold and very narrow, in contrast to quark model expectations. This is what makes it an exotic candidate. Secondly, when quark potential model calculations include the two-meson $DK$ couplings to the $c\bar{s}$ component, the eigenstate mass dramatically shifts downwards and in line with experimental results. Multiple lattice QCD calculations \cite{Cheung:2020mql} have been performed, taking into account the various systematical errors, and find a bound state pole nearby the experimental mass. These lattice QCD results find that both the $c\bar{s}$ and two-meson $DK$ components are important, but the diquark operators are not. For these reasons, the $D_{s0}(2317)$ is a prototype to understanding the $X(3872)$. Yet, the $D_{s0}(2317)$ is easier to study theoretically. We can use our understanding of the $D_{s0}(2317)$ as input into understanding the $X(3872)$.


It is useful to ask \cite{Eichten:2019may} what perturbations can be performed to the $D_{s0}(2317)$ system in order to understand exotics more quantitatively, and the $X(3872)$ specifically? With lattice QCD, we can explore how exotics change (both their mass, pole type, and residue) as we vary the quark mass - a task that experiment cannot perform. In this way, we can supplement experimental data with lattice QCD. Specifically, \cite{Eichten:2019may} advocates examining the $D_{s0}\to D~ K$ process as a function of strange quark mass $m_s$, and using leading order heavy-quark effective field theory to describe heavy-light systems, and leading order chiral perturbation theory for light-light pseudoscalar mesons. If we take $m_s\to  m_s - \epsilon$, then $M_D$ reduces by $-\epsilon$, while $M_K^2$ changes by $-B\epsilon$, with $B$ the leading order $\chi$-PT constant. Applying this logic to $D_{s0}\to D~ K$, we see that we can cause  $D_{s0}$ to sit right at threshold, and even to lie above threshold. Studying how the $D_{s0}$ changes would quantitatively illuminate the role of the S-wave threshold effects on exotic states. It would also be interesting to see the bound state pole move in the complex plane and become a virtual/resonance state as the threshold effects change as a function of $m_s$. We could even move the binding energy to be exactly that of the $X(3872)$, and understand this arch-typical state more quantitatively.

Such a calculation is computationally feasible also. All the expensive pieces can be reused, and only the cheap unphysical strange quark propagators would need to be recomputed. Additionally, the systematic errors would be minimal. Consequently, such a program is imminently needed and has few minimal downsides. 

\section{Conclusion}

Taken together, we have explored the latest  exotic state developments. This has included the experimentally established $X(3872)$, $Y(4230)$, $Z_c(3900)$, $Z_b(10610)$, the recently discovered all charm tetraquark $X(6900)$, and the $D_{s0}(2317)$. We also discussed the non-existence of a bound all-bottom tetraquark, how lattice QCD can be used to study slight perturbation of the $D_{s0}(2317)$ in order to quantitatively understand S-wave threshold effects, and how imminent results are needed from experiments to discover the bound $bb\bar{u}\bar{d}$ state. 

We did not get to discuss all of the interesting exotic states currently in the literature. However, given the physics we have learned, we can generalise our knowledge to a multitude of other states. For example, the $Z_b(10650)$ is likely the $B^*\bar{B}^*$ spin-partner of the $Z_b(10610)$. Similarly, the $Z_c(4020)$ is likely the $D^*\bar{D}^*$ spin-partner of the $Z_c(3900)$. The heavy flavour equivalent of the $X(3872)$, the $X_b$, has likely not been seen as the $\chi_{b1}(2P)$ is below the open-bottom threshold. This is in contrast to the charm case, which highlights the importance of the $\chi_{c1}(2P)$ for the $X(3872)$. The pentaquark $P_c$ is likely a molecule of $\bar{c}c$ and a nucleon. LHCb have released preliminary results about a exotic flavour $cs\bar{u}\bar{d}$ state, where the lattice QCD HadSpec collaboration finds suggestions of this state in the $I=0$, $J^P=0^+$ channel.

There are other established states that all models fit accurately. More experimental data is needed to nullify some of the models describing these states. The exotic candidates in this class are the $Y(4360)$ (which could be the $D_1D^*$ partner of the $Y(4230)$), the $Y(4660)$ (which could be the $D_sD_{s1}$strange partner of the $Y(4230)$), and the $Z(4430)$. 

The future of exotic spectroscopy is bright. With the BELLE and BES  upgrades, the new PANDA experiment, and the continuous measurements from the LHC, further data continues to be collected and analysed. Such work will continue to find more exotic candidates, and further shed light into the exotic ways Nature operates at the smallest of scales.

\bibliographystyle{JHEP}
\bibliography{main}

\end{document}